\documentclass[a4paper,11pt]{article}

\usepackage[utf8]{inputenc}
\usepackage[T1,T2A]{fontenc}
\usepackage{amsmath,amsfonts,amssymb}

 \usepackage{graphicx}

\usepackage{geometry}
\allowdisplaybreaks

\begin{document}

\begin{titlepage}

\begin{flushright}
KANAZAWA-23-01
\end{flushright}

\vspace{1.5cm}

\begin{center} {\LARGE \bf Sterile neutrino dark matter: relativistic freeze--out  } \end{center}

\vspace{1cm}

\begin{center}
  {\bf Oleg Lebedev\(^a\) and   
  Takashi Toma\(^b\)}
\end{center}
  
\begin{center}
  \vspace*{0.15cm}
  \it{\({}^a\)Department of Physics and Helsinki Institute of Physics,\\
  Gustaf H\"allstr\"omin katu 2a, FI-00014 Helsinki, Finland}\\
  \vspace*{0.15cm}
  \it{\({}^b\)Institute of Liberal Arts and Science, Kanazawa University, \\Kanazawa
920-1192, Japan\\
Institute for Theoretical Physics, Kanazawa University, \\Kanazawa
920-1192, Japan}
\end{center}
  
\vspace{2.5cm}

\begin{center} {\bf Abstract} \end{center}
 
 \noindent Long-lived sterile neutrinos can play the role of dark matter. We consider the possibility that such neutrinos form a thermal bath with a singlet scalar, while not being in thermal equilibrium with the Standard Model fields. Eventually, the neutrino dark matter undergoes freeze-out in the dark sector, which can occur in both non-relativistic and relativistic regimes. To account for the latter possibility,
we use the full Fermi-Dirac and Bose-Einstein distribution functions with effective chemical potential in the reaction rate computation. This allows us to 
study the freeze-out process in detail and also obtain the necessary thermalization conditions. We find that   relativistic freeze-out occurs in a relatively small part of the parameter space.
In contrast to the standard weakly-interacting-massive-particle (WIMP) scenario, 
 the allowed  dark matter masses extend to $10^4$ TeV without conflicting perturbativity.

\end{titlepage}

\tableofcontents

\section{Introduction}

The lightest sterile neutrino is among the best motivated candidates for dark matter (DM).  The small but non-zero masses of the active neutrinos strongly suggest the existence of their right-handed counterparts \cite{Minkowski:1977sc,GellMann:1980vs,Yanagida:1979as,Mohapatra:1979ia,Schechter:1980gr,Lazarides:1980nt},
although further details remain unknown. If the latter are very long-lived, they can constitute all of the observed dark matter in the Universe 
\cite{Dodelson:1993je, Shi:1998km,Abazajian:2001nj, Asaka:2005an,Asaka:2006nq}.

There are distinct mechanisms for production of  sterile neutrinos in the early Universe. They can be produced by the Standard Model (SM) thermal bath via the active-sterile mixing
\cite{Dodelson:1993je}, while in singlet extensions of the SM there are further channels due to allowed  scalar-neutrino couplings \cite{Kusenko:2006rh,Petraki:2007gq,Hansen:2017rxr}. In this work, we study the possibility that the lightest sterile neutrino reaches thermal equilibrium with the singlet scalar and
subsequently freezes out. 
Thermal neutrino  freeze-out has been shown to be a viable 
option for obtaining the correct DM relic abundance in various models 
(see e.g.,\;\cite{Herms:2018ajr,Jaramillo:2020dde,Arcadi:2021doo,Coy:2022unt}).

The relevant dynamics of sterile neutrinos can take place in the relativistic regime, where the usual Maxwell-Boltzmann approximation becomes inadequate.
This calls for a fully relativistic approach \cite{Arcadi:2019oxh,Lebedev:2021xey}, which takes into account the quantum statistical effects as well as the effective neutrino chemical potential, necessary to describe the freeze-out process.
To this end, we derive the relativistic reaction rates, include the thermal masses     and solve the Boltzmann equation without resorting to the Maxwell-Boltzmann approximation.

Some aspects of relativistic neutrino dynamics were studied in \cite{DeRomeri:2020wng,Bandyopadhyay:2020ufc,Bringmann:2021sth}, although in a different context, e.g. related to freeze-in dark 
matter.\footnote{Relativistic effects in scalar decay into sterile neutrinos were studied in \cite{Drewes:2015eoa}.} In the present work, the quantum statistics effects are important for both initial and final states of the freeze-out process, which makes the reaction rate calculation significantly more complicated. The dependence on the effective chemical potential also becomes non-trivial in our model.
The main outcome of our analysis is the allowed parameter space in terms of the sterile neutrino mass and coupling, where both relativistic and non-relativistic freeze-out is consistent 
with the observed DM relic density.

\section{The model}

The dark sector of the model consists of a sterile Majorana neutrino $\nu$ and a real scalar $s$. 
It is similar to the model we studied in \cite{DeRomeri:2020wng}, although here we make some simplifications in the scalar potential 
and also allow for a bare Majorana neutrino mass $M$.  
In the 4-component notation, the relevant part of the Lagrangian is 
 \begin{equation}
    -\Delta {\cal L}_\nu = {1\over 2} \lambda\; s \; \bar \nu \nu  + {1\over 2} M \; \bar \nu \nu \;,
  \end{equation}
while dominant part of the  $s$-scalar potential is given by
 \begin{equation}
    -\Delta {\cal L}_s =  {1\over 2} m_s^2 s^2 +
    {1\over 4} \lambda_s s^4  \;.
      \end{equation}
  The couplings of these states to the Standard Model fields are taken  to be feeble.    The sterile neutrino is assumed to be very long-lived such that it can constitute all of the dark matter in the 
  Universe. 
  Similar set-ups have been considered in \cite{Kusenko:2006rh,Petraki:2007gq,Hansen:2017rxr,Flood:2021qhq}, while  more sophisticated constructions can be found in
   \cite{Fernandez-Martinez:2021ypo,Belanger:2021slj,Seto:2020udg,Eijima:2022dec}.

        \begin{figure}[h!] 
\centering{
 \includegraphics[scale=0.78]{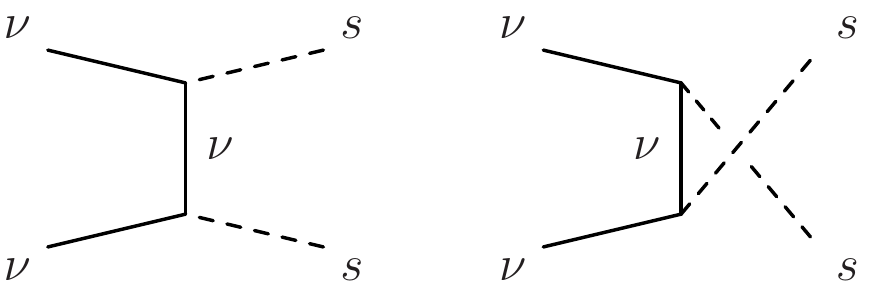}
}
\caption{ \label{diag}
Sterile neutrino annihilation diagrams.
 }
\end{figure}

The evolution of the dark sector is quite simple.
For a sufficiently large $\lambda$, the neutrino-scalar system reaches thermal equilibrium. 
Since there is essentially no communication with the SM, 
the corresponding temperature  $T$ can be very different from the temperature of the visible sector $T_{\rm SM}$.
Eventually, the sterile neutrinos fall out of thermal equilibrium and freeze out via (Fig.\,\ref{diag})
\begin{equation}
 \nu + \nu \rightarrow s+s \;,
 \end{equation}
 as long as $M > m_s$. The freeze-out process can take place in the relativistic or non-relativistic regimes, depending on the $M$ and $\lambda$,
 and produce the correct DM relic abundance.
 Subsequently,  the scalar   decays into the SM states due to its feeble coupling to the observable sector, e.g. via higher dimensional operators.
 If this occurs before the BBN, such a decay causes no damage to light nuclei. The above mechanism of obtaining the required DM relic density via annihilation into unstable states 
 is similar to that of ``secluded'' dark matter \cite{Pospelov:2007mp}.

\begin{figure}[h!] 
\centering{
 \includegraphics[scale=0.58]{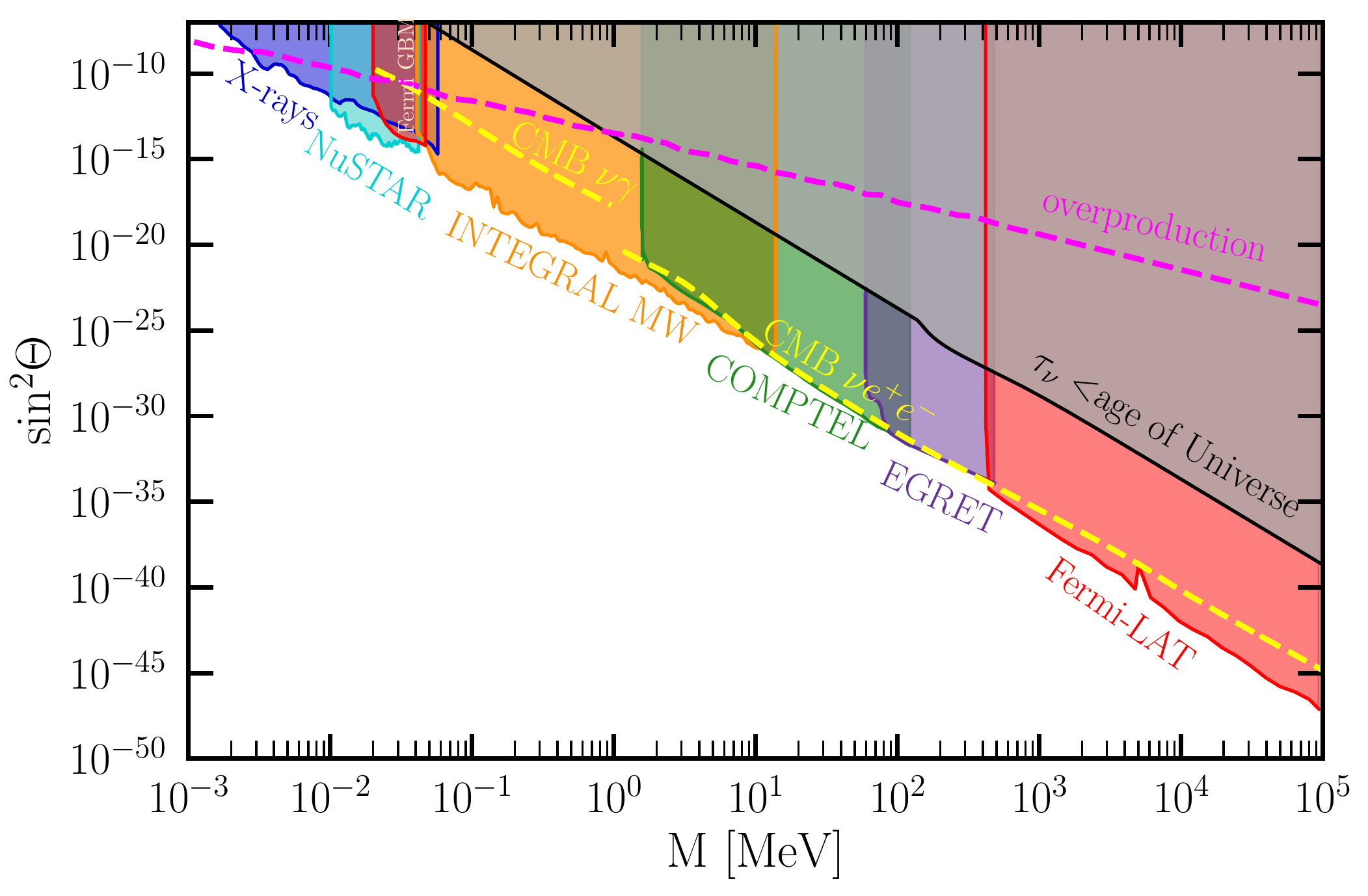}
}
\caption{ \label{limits}
Constraints on the active-sterile neutrino mixing angle $\Theta$ from astrophysics and cosmology. (See \cite{DeRomeri:2020wng} for details.)
 }
\end{figure}

 Throughout this paper we assume that the 
  the active-sterile mixing angle $\Theta  $ is tiny such that the lifetime of the sterile neutrino is much longer than the age of the Universe. The cosmological and astrophysical constraints 
  on $\Theta$ are shown in Fig.\,\ref{limits} (see e.g.,\;\cite{Boyarsky:2009ix,Boyarsky:2018tvu}). This is an update and extension of the analogous figure in Ref.\,\cite{DeRomeri:2020wng}, which contains a detailed description of the bounds.\footnote{We are grateful to Valentina De Romeri for updating the figure.} 
  The allowed mixing has to be extremely small, $\Theta < 10^{-10}$ for $M >1\;$MeV. This may be a result of the $Z_2$ symmetry acting just on the sterile neutrino $\nu \rightarrow -\nu$
 and being broken by Planck-suppressed operators, although the specifics are not important for our purposes.

 We note that the existence of heavier sterile neutrinos which decay into the lightest one does not affect our considerations due to the dark sector thermalization. This is in contrast with the
 non-thermal neutrino DM, e.g. freeze-in neutrinos \cite{DeRomeri:2020wng}, whose abundance is affected by many factors including quantum gravity \cite{Lebedev:2022cic}.   
   The non-zero masses of the active neutrinos can result from  their Yukawa couplings to heavier sterile neutrinos, of which there could be many \cite{Ellis:2007wz}.

 \section{Reaction rate with the Bose--Einstein and Fermi--Dirac statistical factors}

 The main ingredient in our study is the relativistic   $\nu \nu \rightarrow ss$ reaction rate. 
  The $a \rightarrow b$ rate per unit volume is given by \cite{Kolb:1990vq},
\begin{equation}
\Gamma_{a\rightarrow b} = \int \left( \prod_{i\in a} {d^3 {\bf p}_i \over (2 \pi)^3 2E_{i}} f(p_i)\right)~
\left( \prod_{j\in b} {d^3 {\bf p}_j \over (2 \pi)^3 2E_{j}} (1+f(p_j))\right)
\vert {\cal M}_{a\rightarrow b} \vert^2 ~ (2\pi)^4 \delta^4(p_a-p_b) \;,
\label{Gamma}
\end{equation}
with $a$ and $b$ being a fermionic and a bosonic state, respectively.
Here $p_i$ and $p_j$ are the initial and final state momenta, and 
${\cal M}_{a\rightarrow b}$ is the  QFT  $a \rightarrow b$  transition amplitude.
In our convention, we absorb in $\vert {\cal M}_{a\rightarrow b} \vert^2$ both
 the {\it  initial and final} state phase space symmetry factors.
 $f(p)$ is the momentum distribution function, which for scalars (fermions)  in  kinetic equilibrium takes the  Bose-Einstein (Fermi-Dirac) form  
 \cite{Bernstein:1988bw},\cite{Bernstein:1985th},
\begin{equation}
f(p)= {1 \over \exp^{E-\mu\over T} \mp 1 } \;,
\end{equation}
where $\mu$ is the {\it effective} chemical potential accounting for deviation from chemical equilibrium.
The  factor $1+ f(p_j)$ reflects  the Bose enhancement of the reaction rate due to degenerate final states.

 In our model, the scalar has a  significant self--coupling such that it maintains full thermal equilibrium   during the period of interest. 
  On the other hand, the sterile neutrinos fall out of thermal equilibrium in the freeze-out regime, while maintaining kinetic equilibrium with the scalars.\footnote{We note that the reaction
  $\nu\nu \rightarrow ss$ is less efficient than $\nu s \rightarrow \nu s$ is, partly due to $n_s > n_\nu$ after freeze-out.}
  This requires introduction of the neutrino effective chemical potential $\mu$:
 \begin{equation}
 f_\nu(p) =  {1\over \exp^{{E-\mu \over T} }+1} ~~~ , ~~~  f_s(p) = {1\over \exp^{{E \over T} }-1} \;.
 \end{equation}
Note that the $\mu$--contribution does not factorize, unlike it does in the non--relativistic limit.

 Using the approach detailed in Refs.\,\cite{Arcadi:2019oxh,Lebedev:2021xey}, the rate can be expressed as an integral over the thermally-modified cross section in the center-of-mass (CM) frame.
 This generalizes the well-known Gelmini-Gondolo formula \cite{Gondolo:1990dk} to relativistic energies. 
 The difference from our previous fully bosonic result \cite{Arcadi:2019oxh,Lebedev:2019ton} lies in the angular integration over the initial state, which due to the Fermi-Dirac statistics leads to the replacement of the hyperbolic 
   sine with a hyperbolic cosine under the logarithm, such that
 \begin{eqnarray}
&& \Gamma_{\nu\nu\rightarrow ss} =  N^2 \times (2\pi)^{-6} \int d^3 {\bf p_1} d^3 {\bf p_2} ~f_\nu (p_1) f_\nu (p_2) ~\sigma (p_1,p_2) v_{\rm M\o l} = \label{Gamma24} \\
&& N^2 \times {4 T \over \pi^4} \int_M^\infty dE ~E^3 \sqrt{E^2-M^2} \int_0^\infty d\eta {    \sinh \eta \over e^{ 2(E  \cosh\eta -\mu)/T }-1}~
\ln {  \cosh    {E\cosh\eta + \sqrt{E^2 -M^2} \sinh\eta -\mu \over 2T}    \over
\cosh   {E\cosh\eta - \sqrt{E^2 -M^2} \sinh\eta -\mu \over 2T}  } \nonumber \\
&& \times \sigma_{\rm CM}(E,\eta) \;, \nonumber
\end{eqnarray}
 where $M$ is the sterile neutrino mass, $E$ is the particle energy in the CM frame, $\eta$ is the rapidity,  and  $N = 2$ accounts for the spin degrees of freedom, i.e. the single-particle phase space element is $d\Pi =
 N \times  (2\pi)^{-3 } d^3{\bf p} /(2E)$.
 In our convention, the symmetry factors due to the identical particles in the initial and final states, namely, $1/(2!2!)$,
have been absorbed into $\sigma$.
The thermally-modified $\nu\nu\rightarrow ss$ cross section in a general frame is defined by 
\begin{equation}
\sigma (p_1,p_2)= {1\over 4 F(p_1,p_2)} \int \vert {\cal M} \vert^2 (2\pi)^4 \delta^4 \left(p_1+p_2 -\sum_i k_i\right)
\prod_i {d^3 {\bf k}_i \over (2 \pi)^3 2E_{k_i}} \left(1+f_s(k_i)\right) \;,
\label{sigma-def}
\end{equation}
with $F=\sqrt{(p_1 \cdot p_2)^2 - M^4} $ and $\vert {\cal M} \vert^2$ including the symmetry factor  $1/(2!2!)$ as well as the usual spin averaging and summation.
 The 
  M\o ller velocity appearing in the first line of  (\ref{Gamma24})  is given  by 
\begin{equation}
v_{\rm M\o l}= {F(p_1,p_2) \over E_1 E_2} \equiv { \sqrt{(p_1 \cdot p_2)^2 - M^4} \over E_1 E_2 } \;.
\end{equation}
Note that the dependence on the scalar mass $m_s$ appears only in $\sigma (p_1,p_2)$.

The cross section can be computed numerically in the CM frame ($ \sigma_{\rm CM}$), where 
the final state quantum statistical factors take the form
\begin{equation}
1+f_s(k_i)= 1+ {    1  \over e^{ (k_i^0 \cosh \eta + k_i^3 \sinh \eta)/T}  -1   } \;,
\end{equation}
 with $k_{1,2}$ being  the final state particle 4-momenta in the CM frame.  
 The amplitude  involving fermions has a non-trivial angular dependence, which makes the analysis complicated.
 Numerically, 
 the cross section calculation can, for example, be done with  the help of CalcHEP \cite{Belyaev:2012qa} by absorbing $1+f(k_i)$ into a 
 momentum-dependent vertex.

  The rate of the inverse reaction $ss \rightarrow \nu  \nu$ is also needed for the analysis.
  It contains the combination $f_s (k_1) f_s (k_2) ( 1- f_\nu(p_1)) (1- f_\nu (p_2))$, which yields  
 \begin{equation}
 \Gamma_{\nu\nu \rightarrow ss} =   \Gamma_{ss \rightarrow \nu\nu} \; e^{2\mu /T} \;.
 \end{equation}
 Therefore, it does not have to be computed independently.

 It is important to note that the $\nu\nu \rightarrow ss$ amplitude is velocity-suppressed in the non-relativistic limit as follows from $C,P$ conservation. This implies, in particular, that 
 indirect detection of neutrino DM would be challenging.

 \subsection{Thermal masses}

 At high temperatures, the mass parameters appearing in $\Gamma_{\nu\nu\rightarrow ss}$ receive important thermal corrections:
 \begin{equation}
  \Delta m_s^2 = {\lambda_s \over 4} \, T^2  ~~,~~  \Delta M^2 =  {\lambda^2 T^2 \over 16} ~,
  \end{equation}
 where we have neglected   ${\cal O}(\lambda^2)$ contributions to $m_s^2$.
 These corrections are necessary for the right high temperature behaviour of the rates \cite{Arcadi:2019oxh}.

  \begin{figure}[h!] 
\centering{
 \includegraphics[scale=0.74]{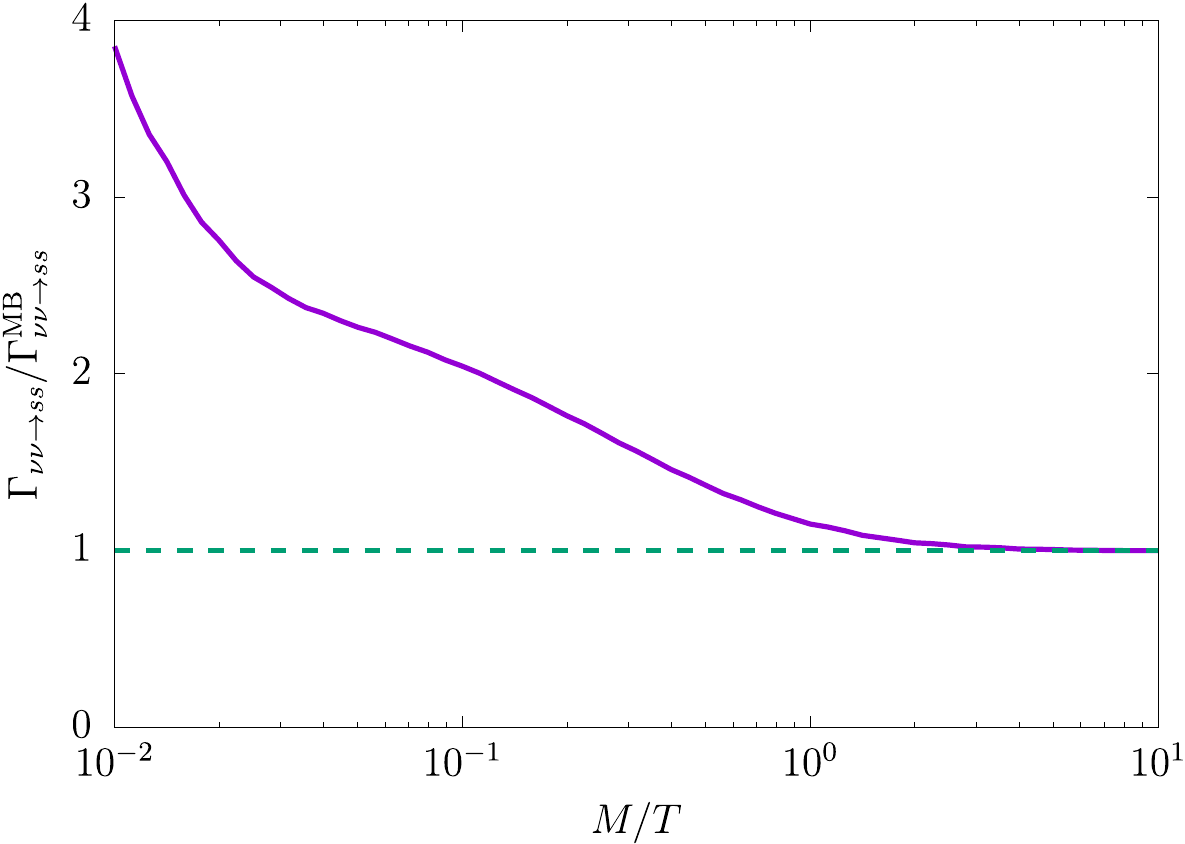}
}
\caption{ \label{q}
The effect of quantum statistics on the  reaction rate of ${\nu\nu \rightarrow ss}$: the curve represents the ratio of $\Gamma_{\nu\nu \rightarrow ss}$ and its Maxwell-Boltzmann limit
$\Gamma_{\nu\nu \rightarrow ss}^{\rm MB}$ at $\mu=0, \, \lambda \ll 1$. 
 }
\end{figure}

 Replacing the bare masses  in the reaction rate with the thermally corrected ones, we obtain the final expression which we use in our numerical analysis.
It is instructive 
   to evaluate the impact of quantum statistics on $\Gamma_{\nu\nu \rightarrow ss} $. Our result is presented in Fig.\,\ref{q}.  At $T \sim M$, 
  quantum statistics accounts for about 20\% correction to the reaction rate, while at $T \sim 10^2 \,M$ the effect increases to a factor of 4. 
  The Fermi-Dirac distribution for the neutrinos  leads to a mild suppression of the rate, while its Bose-Einstein counterpart for the scalars entails a significant increase in the rate such that the 
  net result is an overall enhancement of $\Gamma_{\nu\nu \rightarrow ss}$.  
  
  At $T \sim M$, which is the most relevant regime in our work, the effect of thermal masses is only modest.
 We note that at very high temperatures, $T \gg M$, the thermal mass contributions make the 
   reaction $s \leftrightarrow \nu\nu$   kinematically allowed for $\lambda_s \gtrsim \lambda^2$ even if the scalar is very light at zero temperature. However, we are interested in $\lambda_s \ll 1$, e.g. $\lambda_s \sim 10^{-2}$, which 
   forbids this reaction for $T < 10M$ and makes it irrelevant to our analysis.

  \section{Thermodynamic evolution and dark matter abundance}
  
Our general expression for the reaction rates allows us to study the evolution of the system in both relativistic and non-relativistic regimes, and compute the resulting
abundance of dark matter.

  \subsection{Boltzmann equation}
  
  In our model, the scalar is lighter than the sterile neutrino and assumed to have a sufficiently large self-coupling so that
  it remains in thermal equilibrium
   during  the period of interest, in particular, after neutrino freeze-out.
    Thus, the only two thermodynamic variables in the dark sector  are 
  the neutrino chemical potential $\mu$, which governs $n_\nu$ throughout freeze-out,  and the dark temperature $T$.   
 
 The evolution of $\mu$ and $T$ is dictated by 
   the Boltzmann equation for the neutrino number density and     the dark entropy conservation constraint,
  \begin{eqnarray}
 && {d n_\nu \over dt} + 3H n_\nu = 2 \left(  \Gamma_{ss \rightarrow \nu\nu} -  \Gamma_{\nu\nu \rightarrow ss} \right) \;, \nonumber \\
 && {s_\nu + s_s \over s_{\rm SM}} = {\rm const } \;,
 \label{eqs}
  \end{eqnarray}
  where $s_{\rm SM}$ is the SM entropy density. The entropy density of the dark sector fields  is defined by 
   \begin{equation}
  s= {\rho + p -\mu n \over T}  \;,
  \end{equation}
  where for the neutrinos we have 
  \begin{equation}
  n = N \int {f (p)   } {d^3 {\bf p} \over (2\pi)^3} ~~,~~ \rho = N \int { E \,f (p) } {d^3 {\bf p} \over (2\pi)^3}~~,~~
   p = N \int { | {\bf p}|^2 f(p) \over 3E}\, {d^3 {\bf p} \over (2\pi)^3} \;,
  \end{equation}
  with $N=2$ counting the spin degrees of freedom. These expressions apply to the scalar with $N=1$. 
  Since the system evolves through a semi-relativistic regime, it is important to use the integral form of the thermodynamic quantities instead of 
  approximate asymptotic expressions.

  Solving the system (\ref{eqs}) gives us $\mu (t)$ and $T(t)$, which determine $n_\nu (t)$. The time dependence can be traded for the $T_{\rm SM}$ 
  dependence since $T_{\rm SM} (t)$ is determined entirely by the SM sector entropy conservation. Here we assume that the dark sector
  is cooler than the SM thermal bath such that it does not significantly affect the total energy density and the expansion rate.
  
  The integro-differential system of equations (\ref{eqs}) is quite complicated, in particular, due to the non-factorizable $\mu$-dependence of the rates.
  To solve it numerically, we use the CalcHEP package \cite{Belyaev:2012qa} to compute the thermally-modified cross section
 for   $\nu\nu \rightarrow ss$, which includes the Bose-Einstein final state enhancement factors into the phase space integration.
 Feeding this output into the integral for the rates, we obtain a pointwise solution for $\mu (t)$ and $T(t)$.
 This allows us to study the freeze-out process in detail, without resorting to the non-relativistic approximation.

  \subsection{Thermalization}

     \begin{figure}[h!] 
\centering{
 \includegraphics[scale=0.90]{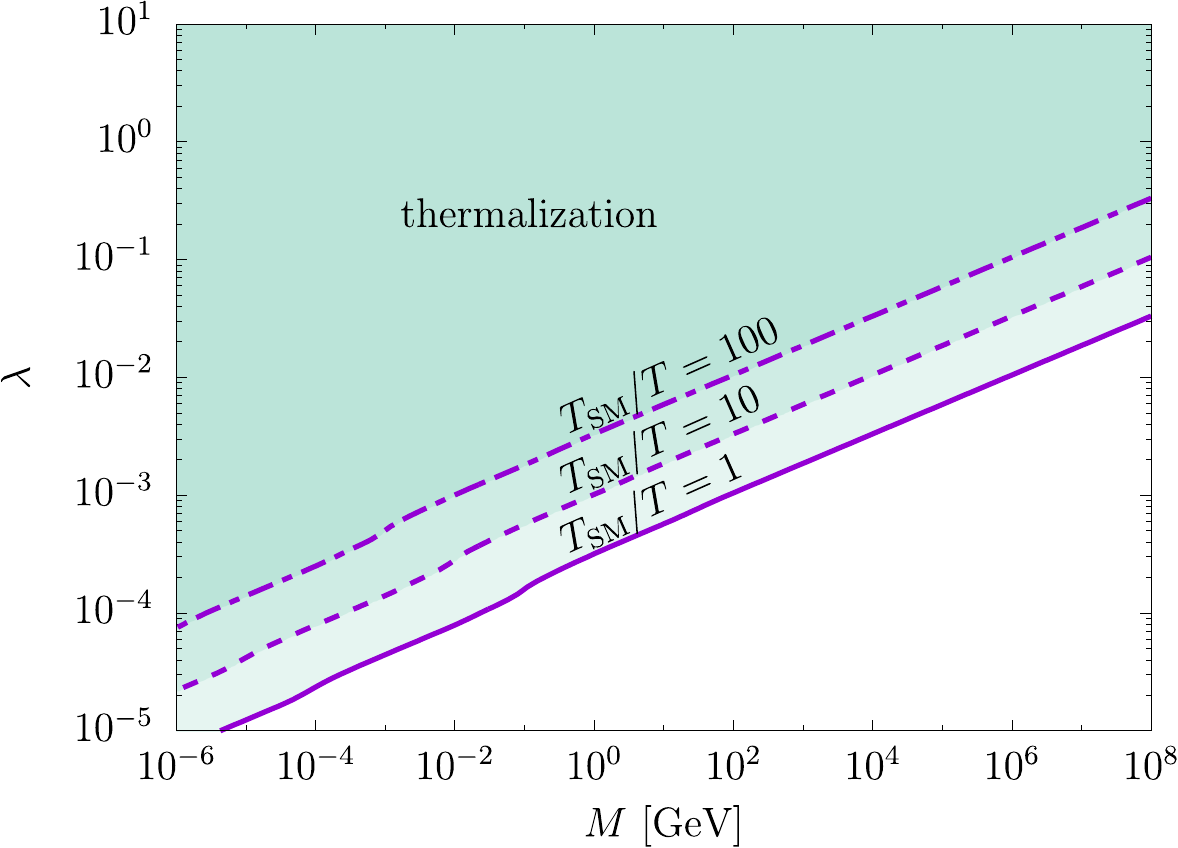}
}
\caption{ \label{therm}
Thermalization constraints due to $\nu \nu \leftrightarrow ss$ for $M \gg m_s \, , \mu=0$ and different temperature ratios $T_{\rm SM}/T$ in 
the relativistic regime.
 }
\end{figure}

  The neutrino-scalar system is assumed to be in thermal equilibrium initially. This is only possible for sufficiently  large $\lambda$-couplings.
  The necessary condition for thermalization is that there exists a temperature  at which the following inequality is satisfied:
   \begin{equation}
  \Gamma_{\nu\nu \leftrightarrow ss} \gtrsim n_\nu H \;,
  \label{thermaliz-cond}
  \end{equation}
i.e. the reaction rate is greater than the expansion rate.
  We find that   the ratio $\Gamma/(n_\nu H)$ is maximized at $M/T =0.6$ for a fixed $T_{\rm SM}/T$. 
  Our assumption is that the energy density of the Universe is dominated by the SM contribution, hence we focus  on 
``cool'' dark sectors,   $T_{\rm SM}/T \geq 1$. 
  The consequent lower bounds on $\lambda$  for different $T_{\rm SM}/T$ are shown 
    in Fig.\,\ref{therm}.

We find numerically that the thermalization constraint can be approximated by
  \begin{equation}
  \lambda \simeq 2\times 10^{-4}\;        \left( {T_{\rm SM}\over T} \right)^{1/2} \,           \left( {M\over {\rm GeV}} \right)^{1/4} \;.
  \end{equation} 
Qualitatively, this behavior can be understood as follows: in the relativistic regime, the reaction rate per unit volume  scales as $\lambda^4 T^4$. Then, condition (\ref{thermaliz-cond}) yields
  $\lambda \propto (T_{\rm SM}/ T)^{1/2} T^{1/4}$ and, since in the regime of interest $T\sim M$, we obtain the above scaling.
  
  In what follows, we only consider parameter space consistent with the thermalization assumption, i.e. the green area in Fig.\,\ref{therm}. 
  This distinguishes our analysis from previous work \cite{Konig:2016dzg}, which has focused on the keV scale non-thermal sterile neutrinos. As we will show, such light neutrinos are not allowed in our 
  framework.

    \subsection{Freeze-out}

  As the temperature of the dark sector decreases, the reaction rates drop and eventually the sterile neutrinos fall out of thermal equilibrium, which signifies freeze-out.
   In our model, this  can occur in both non-relativistic and relativistic regimes. If the freeze-out temperature $T_f$ is close to $M$,
   or more specifically $T_f > M/3$,  we call this regime relativistic. Naturally, this occurs at low enough  $\lambda$ couplings.

\begin{figure}[h!] 
\centering{
 \includegraphics[scale=0.64]{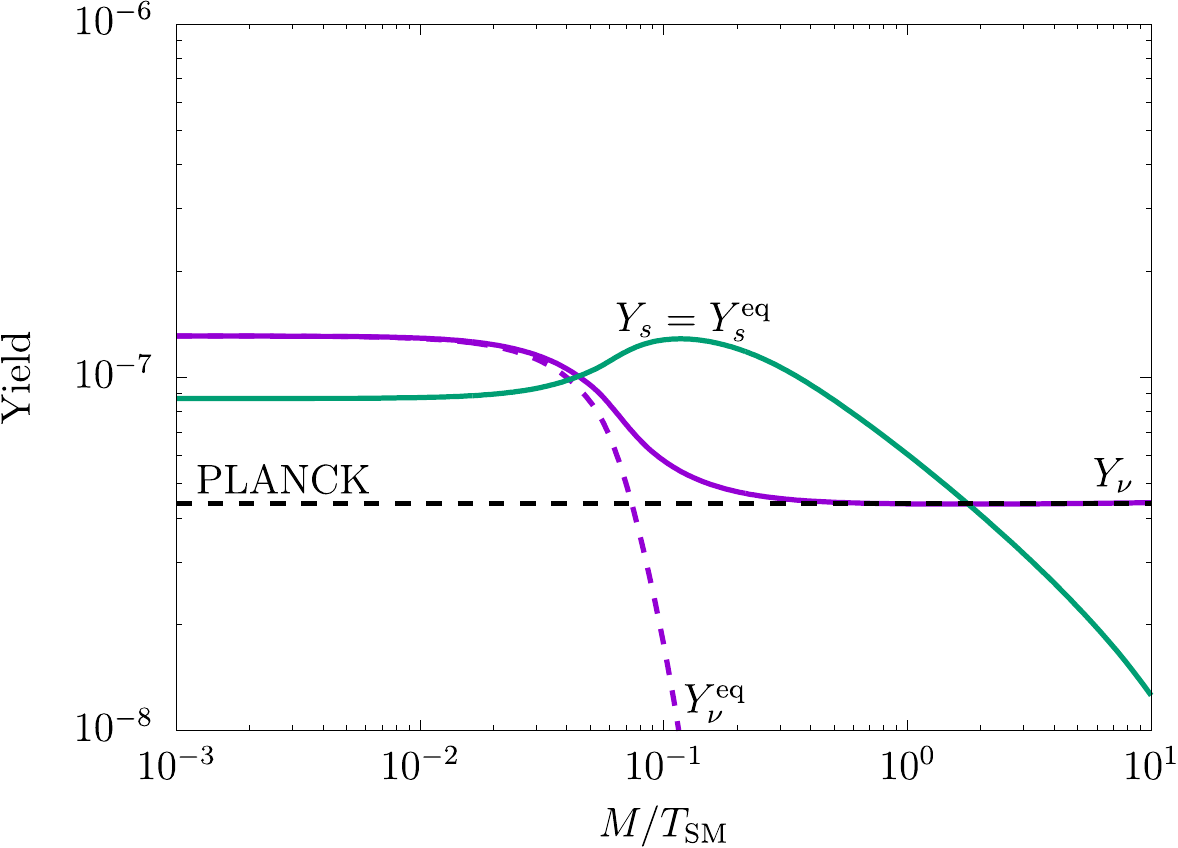}
  \includegraphics[scale=0.64]{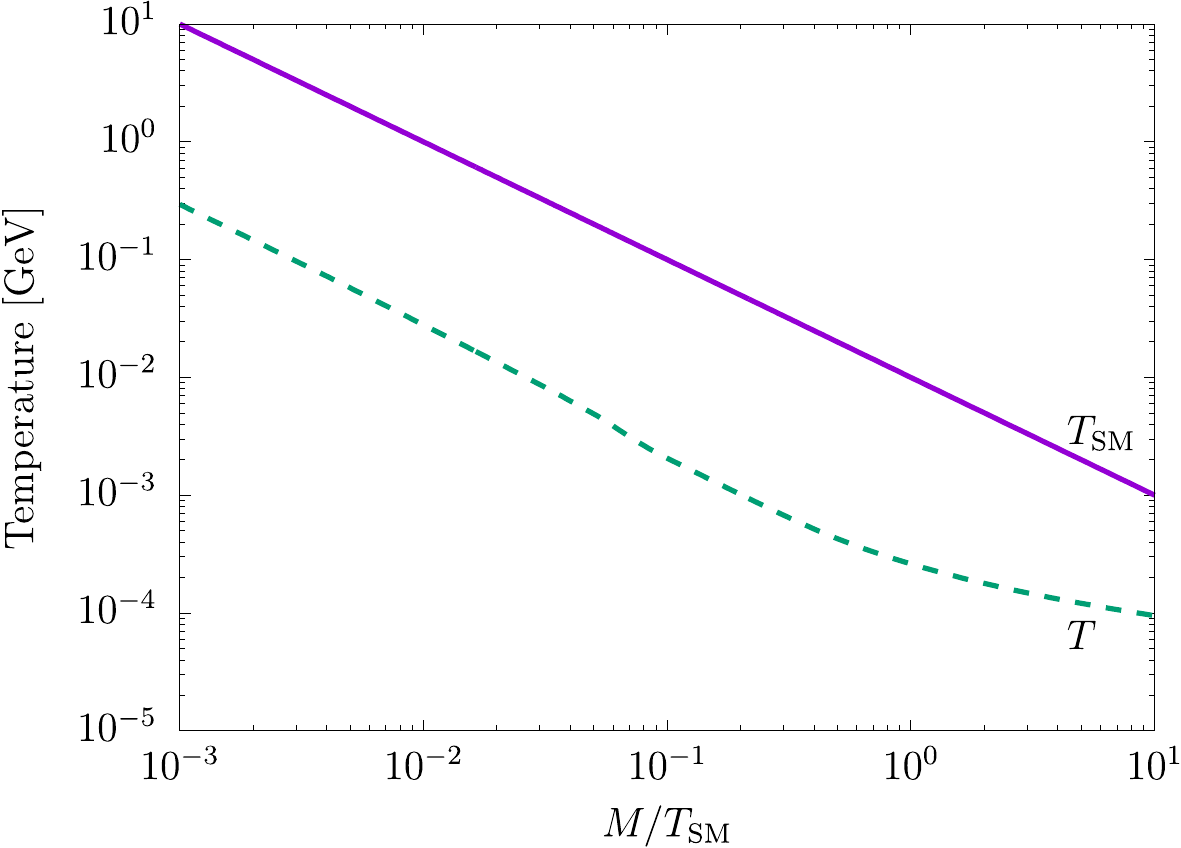}
   \includegraphics[scale=0.64]{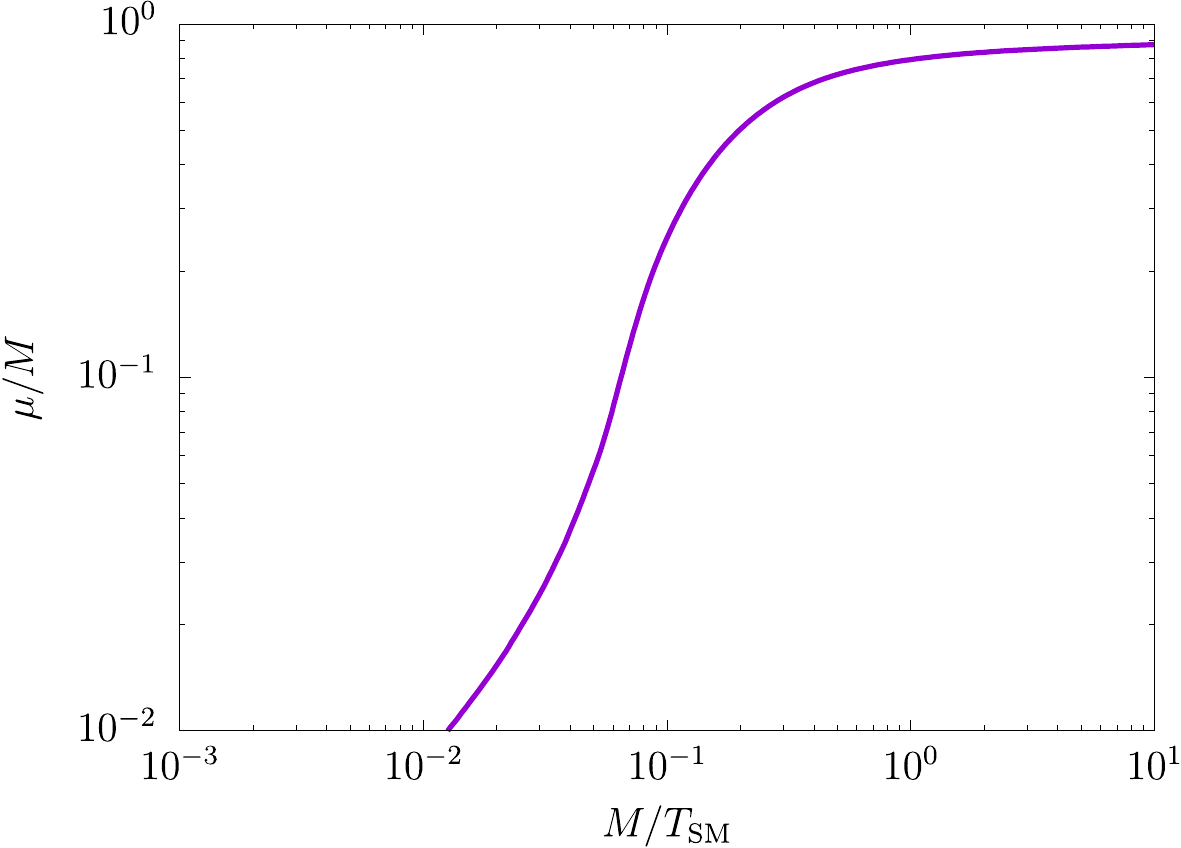}
   \includegraphics[scale=0.64]{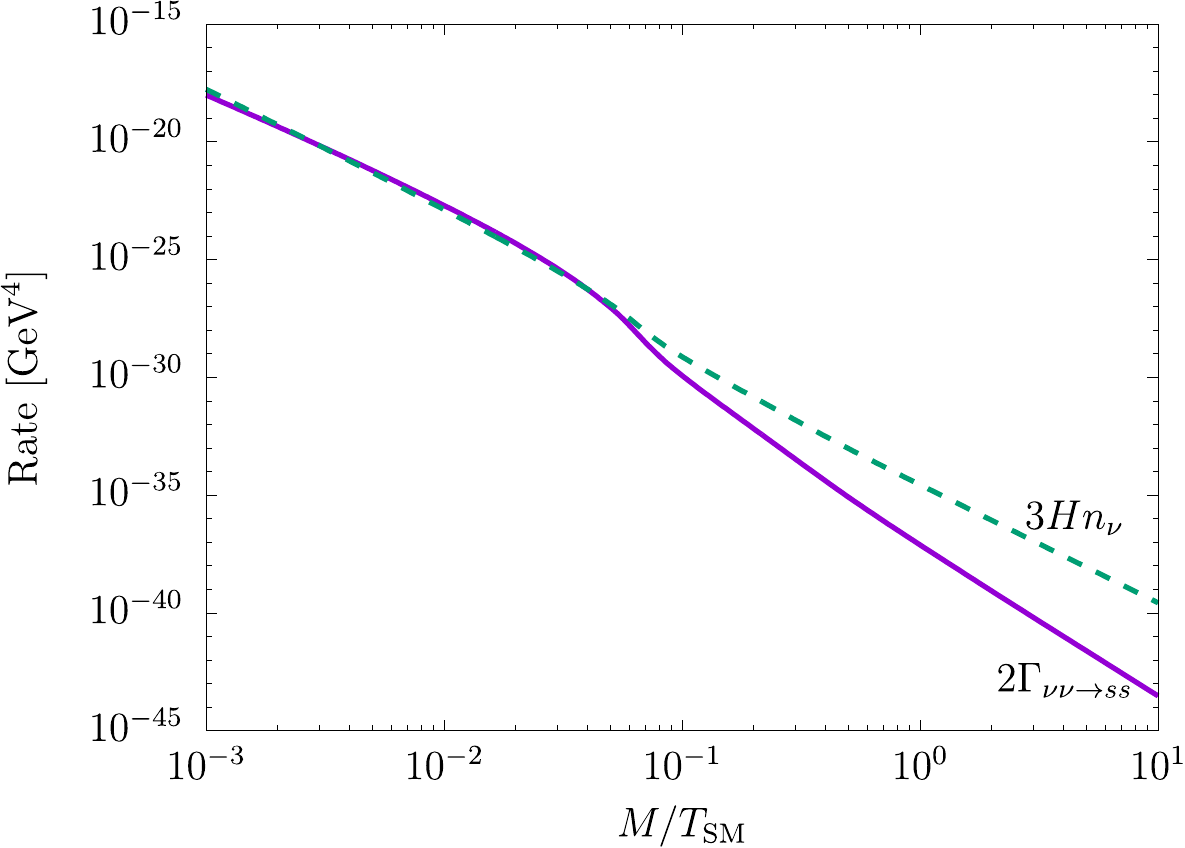} 
}
\caption{ \label{example}
Evolution of the thermodynamic quantities in the relativistic freeze-out regime, $M/T_f =1.54$.
 The parameters are fixed at  $M=10$ MeV, $\lambda=6.5\times 10^{-4}$, and the initial condition for the dark temperature is $T_{\rm SM}/T =34$ at $M/T_{\rm SM}=10^{-3}$.
The PLANCK dashed line represents the observed abundance of dark matter.  }
\end{figure}

  Fig.\,\ref{example} shows an example of relativistic freeze-out. At the freeze-out point, $T_f = M/1.54$ such that the neutrinos remain moderately relativistic. 
  The relevant parameters are set to $M=10$ MeV, $\lambda=6.5\times 10^{-4}$, and the initial condition for the dark temperature is $T_{\rm SM}/T =34$ at $M/T_{\rm SM}=10^{-3}$.
Although the scalar sector parameters do not directly affect the computation of the DM relic abundance, they have to be chosen consistent with our assumptions. 
We take for definiteness $\lambda_s =10^{-2}$ and $m_s/M=0.1$. This makes sure that $s$ remains in thermal equilibrium below the $\nu$--freeze-out temperature \cite{Arcadi:2019oxh} and that $s$ is light enough as
  to provide an annihilation channel for the neutrinos.

   The particle abundance is conveniently characterized by 
   \begin{equation}
   Y_i = {n_i \over s_{\rm SM}} ~~,~~ s_{\rm SM} ={2\pi^2 \over 45} \, g_{*s} \, T_{\rm SM}^3 \;,
   \end{equation}
   where $n_i$ is the number  density of the $i$'th species, $s_{\rm SM}$ is the Standard Model entropy  density  at temperature $T_{\rm SM}$ and 
    $g_{*s}$ is the effective number   of SM degrees {} of freedom contributing {} to the entropy.

   Fig.\,\ref{example} shows the evolution of $Y_\nu $ and $ Y_s$.
  Initially, the particle abundances satisfy $Y_\nu =3/2 \, Y_s$ since $\nu$ has 2 degrees of freedom and $n_\nu$ contains a factor of 3/4 associated with the Fermi-Dirac statistics.  
  We observe that $Y_\nu$ starts deviating from its equilibrium value when the $\nu \nu \rightarrow ss$ reaction rate drops below the expansion rate. The neutrino annihilation becomes more efficient than the inverse reaction, hence the number density of $s$ increases while still remaining at the equilibrium value. The scalar sector  ``heats up'' somewhat,  although this is not easily seen 
  in the temperature evolution panel of 
   Fig.\,\ref{example}
  due to a parallel  increase in $T_{\rm SM}$ at around 100 MeV associated with the QCD phase transition.
  
  At this stage, a significant neutrino chemical potential starts to develop. It asymptotically approaches $M$ in the non-relativistic regime, as follows from the number and entropy conservation \cite{Arcadi:2019oxh} .
  The annihilation remains effective for some time after the freeze-out point, which we define as the point at which 
  $Y_\nu$ and $Y_\nu^{\rm eq}$ start to differ (or, equivalently, the reaction rate drops below the expansion rate).  Therefore $Y_f > Y_\infty$ and, as seen in Fig.\,\ref{example},  the eventual DM abundance 
  $Y_\infty$
  can be evaluated via the equilibrium number density 
  $n_\nu^{\rm eq}$ at some later point $\tilde x_f > x_f$,
  where $x_f = M /T_f$ and $\tilde x_f = M /\tilde T_f$. Numerically, we find that $ \tilde{x}_f $ for different parameter choices is  fitted  by
  \begin{equation}
 \tilde{x}_f / x_f = 2.6\, x_f^{-1.04} - 3.0 \,x_f^{-0.024} + 3.6 \;,
 \end{equation}
 such that
  \begin{equation}
  Y_\infty = {n_\nu^{\rm eq} (\tilde T_f) \over s_{\rm SM} (\tilde T_f^{\rm SM}) }\;, 
  \end{equation}
  where $\tilde T_f^{\rm SM}$ 
  is the SM temperature corresponding to the dark sector temperature $\tilde T_f$.
 In the non-relativistic regime, $x_f \sim 10$, the reaction rate drops  exponentially after freeze-out such that  $\tilde x_f \simeq x_f$. On the other hand, in the case of relativistic freeze-out, 
 $x_f \sim 1-2$, neutrino annihilation remains significant for some time and $\tilde x_f  \simeq 2 x_f $, which leads to a large reduction of the neutrino abundance after freeze-out (Fig.\,\ref{example}).
 The observational constraint on $Y_\infty$ is given by \cite{Ade:2015xua}
   \begin{equation}
    Y_\infty =  4.4 \times 10^{-10} \; \left( {{\rm GeV}\over M} \right) \;,
    \end{equation} 
which is represented in Fig.\,\ref{example} by the black dashed line marked ``PLANCK''.

  When $s$ becomes non-relativistic while remaining in thermal equilibrium, the scalar sector starts ``heating up'', i.e. $T$ begins to decrease in time slower than $T_{\rm SM} $ does. This is a known phenomenon
  following from entropy conservation \cite{Carlson:1992fn}. 
  Shortly after $Y_\nu$ reaches a plateaux, $s$ decays into the SM fields.  Since $T < T_{\rm SM}$ and the SM has many more degrees of freedom, this decay does not affect the SM bath in any significant way
  as long as it occurs at $T_{\rm SM} > 1$ MeV.

 \subsection{$s$ decay}
 
 The scalar $s$ is assumed to be unstable such that it decays into the SM states.
 The lifetime of the singlet is chosen in such a way that it decays after $Y_\nu$ reaches its terminal value   $Y_\infty$,  following  neutrino freeze-out. On the other hand, it has to decay before the BBN
 ($T_{\rm SM} \gtrsim 1$ MeV)
 in order not to affect the light nuclei abundance.  The specific mechanism which induces $s$-decay is unimportant for our purposes.

 To give an example, 
 the decay may be triggered by higher dimensional operators, e.g.
 \begin{equation}
 {s\over \Lambda^{4n-3}} \, (F_{\mu\nu} F^{\mu\nu})^n ~~,~~ {s\over \Lambda^{4m-3}} \, (H\bar f_L f_R)^m ~~,~~ {\rm etc.} \;,
 \end{equation}
 where $\Lambda$ is the cuf-off scale of our effective theory. Choosing the  scale and the corresponding Wilson coefficient, one can adjust the $s$-lifetime. 
 For example, if the dominant coupling is that to photons with $n=1$, i.e. $s F_{\mu\nu} F^{\mu\nu}$, the decay width is given by
\begin{equation}
\Gamma_{s\rightarrow \gamma \gamma} = {m_s^3 \over 4 \pi \Lambda^2} \;.
\end{equation}
 Requiring $\tau = \Gamma^{-1}_{s\rightarrow \gamma \gamma} \lesssim 1\;$sec translates into $m_s (m_s/\Lambda)^2 \gtrsim 10^{-23} \;$GeV. For the benchmark value $m_s=1\;$GeV,
 the new physics scale is bounded by $\Lambda \lesssim 0.3\times 10^{12}\;$GeV.

 Another possibility is to allow for a singlet-Higgs mixing, which has a similar effect.
 On general grounds, one expects a linear coupling between the singlet and the Higgs bilinear, 
 \begin{equation}
 \Delta V = {1\over 2} \sigma_{sh} s H^\dagger H \;.
 \end{equation}
 When the Higgs develops a VEV, this interaction leads to a Higgs-singlet mixing, which is responsible for the singlet decay into the SM states. Adjusting the mixing angle $\theta$,
 one obtains the required lifetime. The decay width of the singlet  is obtained by rescaling that of the SM Higgs with the same mass,
  \begin{equation}
 \Gamma_s = \sin^2 \theta ~\Gamma_h \Bigl\vert_{m_h =m_s} \;.
 \end{equation}
The latter is well known, see e.g. Fig.\,28 of \cite{Lebedev:2021xey}. For instance, $m_s=1\;$GeV requires $\theta \gtrsim 10^{-9}$ to obtain the $s$-lifetime below 1 sec.

 The second  option is more constrained since it fixes the relative strength of the SM  couplings. Light scalars mixing with the Higgs are subject to  meson decay constraints\footnote{These constraints can be read off from the analysis of \cite{Andreas:2010ms}, although nominally it has been performed for a light pseudoscalar. The scalar constraints are very similar. } and
 astrophysical bounds \cite{Fradette:2018hhl}. Hence, further constraints may apply to the parameter space of the model, depending on the scalar mass.
 These are evaded if the first option is realized, with an appropriate choice of the relevant operators. The constraints are particularly loose for $s$-decay into multi-particle final states.

   \subsection{Allowed parameter space}

\begin{figure}[h!] 
\centering{
 \includegraphics[scale=0.99]{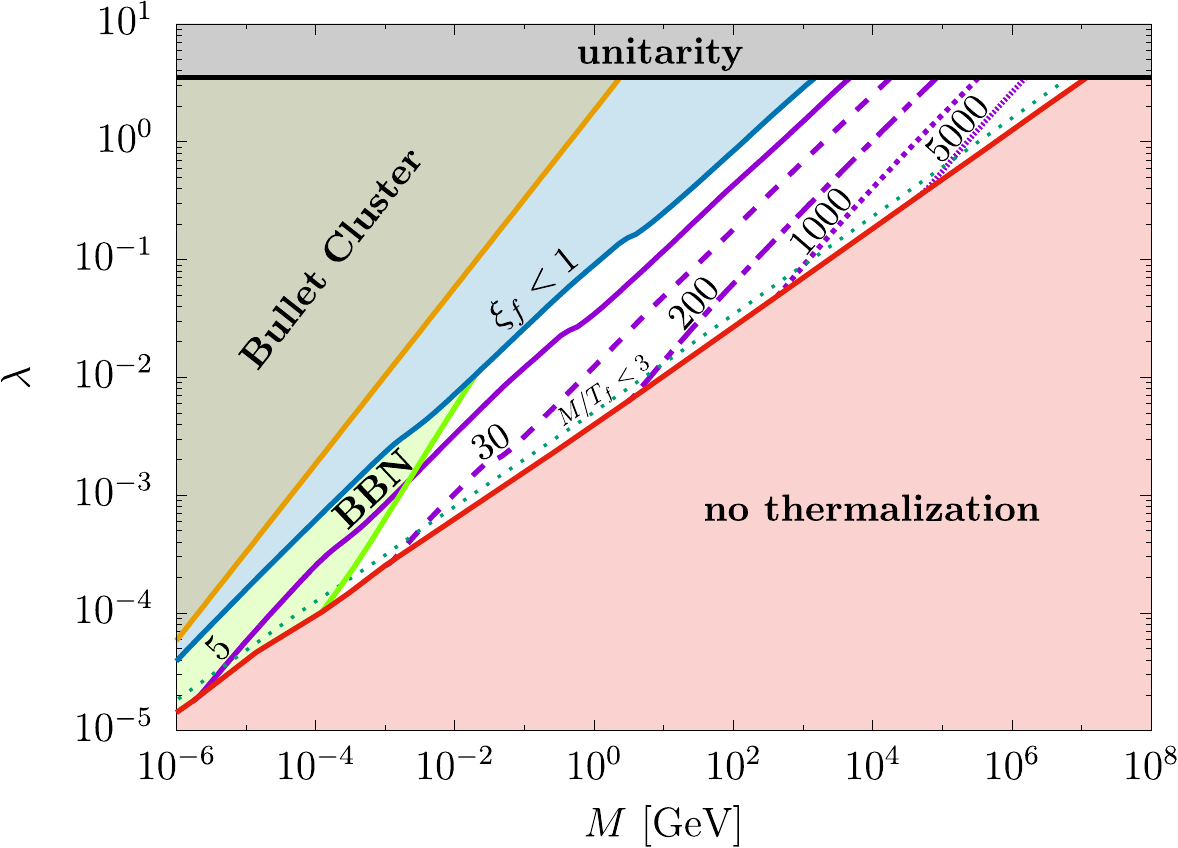}
}
\caption{ \label{par-space}
Allowed $(M,\lambda)$ parameter space. The white area is consistent with all the constraints. The different curves within it correspond to the observed dark matter relic density for a fixed  
 $T_{\rm SM}/T$ ratio at freeze-out, which labels the curves. Below the green dotted line, freeze-out occurs in the relativistic regime, $T_f > M/3$.
 The Bullet Cluster bound is obtained for $m_s=0.1 M$.}
\end{figure}

   The allowed $(M,\lambda)$ parameter space consistent with our assumptions 
   and leading to the correct $cold$ DM relic density 
   is shown in Fig.\,\ref{par-space}. The lower right corner is excluded due to non-thermalization of the $\nu-s$ system.
  Large couplings $\lambda \gtrsim \sqrt{4\pi}$   (grey area)  are inconsistent with perturbative unitarity. 
  
   In the upper left corner, the sterile neutrinos have significant self-interaction which is disfavored by the Bullet Cluster observations (see, e.g.\;\cite{Robertson:2016xjh}).
   In the non-relativistic limit, the neutrino self-interaction is mediated by the $t$- and $u$-channel exchange of the scalar. This leads to the neutrino elastic scattering cross section
   \begin{equation}
   \sigma_{\rm self} \simeq {\lambda^4 \over 8\pi} \, {M^2 \over m_s^4} \;.
   \end{equation}
   The exclusion limit is obtained under the assumption $m_s = 0.1 M$, as before, requiring $ \sigma_{\rm self} /M < 1\, {\rm cm}^2 / {\rm g}$.

   The lower left corner is excluded by the nucleosynthesis (BBN) constraints: in this case, $T_{\rm SM}$ at freeze-out is below 1 MeV such that subsequent  $s$-decay into the SM states
   would grossly affect the light nuclei abundance.
   
   We also exclude the blue area marked by ``$\xi_f <1$'', which corresponds to
   $T> T_{\rm SM}$ at freeze-out ($\xi_f \equiv T_{\rm SM}/T \,\vert_{\rm freeze-out}$) and, hence, our assumption that the Universe energy density is dominated by the SM contribution is violated. 
   In a more general setting, however, some of this parameter space may be recovered.

The upper boundary of the relativistic freeze-out region is marked by the green dotted line. In the model at hand, this region is rather small and extends along the border of the ``non-thermalization''
area.
This can be understood qualitatively as follows: smaller couplings lead to earlier freeze-out (since $\Gamma \sim nH$) and, thus, more significant relativistic  effects for a fixed $M$.
On the other hand, thermalization is more difficult to achieve at such couplings, so that relativistic freeze-out turns out to be only barely compatible 
with thermal equilibrium.
 In contrast, in the scalar dark matter model \cite{Arcadi:2019oxh}, relativistic freeze-out dominates the allowed parameter space due to a much stronger Bose-Einstein enhancement of the rates.

  In the  deep non-relativistic regime, the main parameter dependence of the correct relic density curves is captured by
  \begin{equation}
  \lambda \propto M^{1/2} \, \xi_f^{-1/4} \;,
  \label{relic-curve}
  \end{equation}
  This follows from
  the freeze-out relation $n_\nu \sim H/\langle \sigma v \rangle$.
  More precisely, the right hand side of the above equation should be multiplied by $x_f^{1/2}$, which varies rather slowly in the non-relativistic regime.\footnote{Part of this $x_f$-dependence comes from the velocity suppression of the neutrino annihilation cross section.}
   For relativistic freeze-out, the relic density lines tend to  curve down towards vertical lines since in this case 
  $Y \propto T^3/T_{\rm SM}^3$ becomes almost independent of $\lambda$. 
  
      In the allowed region, dark matter is cold: for very light $\nu$, $M \sim 100\;$keV,  the dark sector is significantly cooler than the SM thermal bath, while for heavier  DM (above 10 MeV), the dark temperature is comparable to or lower than $T_{\rm SM}$. Therefore, in either case DM is non-relativistic at the structure formation temperature around 1 keV.

   We conclude that  viable sterile neutrino masses in our model range from about  ${\cal O}(100)$ keV to as much as 
   $10^4$ TeV, without violating perturbativity.
    This in contrast to the usual WIMP mass upper bound of order 
   10 TeV imposed by perturbative unitarity. The difference stems from the fact that the dark sector can be much colder than the observable one, which suppresses the necessary $\lambda$ (see Eq.\,\ref{relic-curve}). This was also observed in  \cite{Coy:2021ann} in a more general context.
   In the scalar DM model \cite{Arcadi:2019oxh}, however, an analogous effect is not as pronounced due to a different $M, \xi_f$ dependence and the upper bound on the DM mass is of order 100 TeV.

      Finally, let us comment on observational prospects of sterile neutrino DM. Indirect detection is hindered by the velocity suppression of $\nu\nu \rightarrow ss$, which reduces the corresponding cross section by  6 orders of magnitude (see, however, \cite{Johnson:2019hsm}) and makes this detection mode unlikely. Direct detection prospects are also dim, although they  depend  on the $s$-decay mechanism. If $s$ does not have any significant coupling to fermions, the direct detection
      cross section is negligibly small. Otherwise, it is suppressed by at least $\lambda^2 v^2/\Lambda^2$ or $\lambda^2 \sin\theta^2$, where $\Lambda$ is the effective theory cut-off and $\theta$ is a possible singlet-Higgs mixing angle.
      On the other hand, the sterile neutrino decay can produce an observable signal in a wide range of photon frequencies, whose intensity depends on the active-sterile neutrino mixing angle (Fig.\,\ref{limits}).

   \section{Conclusion}
   
   We have studied thermal sterile neutrinos as dark matter candidates in a singlet-extended Standard Model. If the dark sector is cooler than the SM thermal bath, the freeze out mechanism can account for the correct DM abundance,
be it relativistic or non-relativistic freeze out.   
 In order to study the relativistic regime, we have obtained the reaction rates which include the Fermi-Dirac and Bose-Einstein quantum statistical factors as well as dependence on the effective chemical potential.
 This allows us to analyze the freeze-out process in detail as well as obtain the necessary thermalization condition. 
  Our main results are presented in  Fig.\,\ref{par-space}. We conclude that the relativistic freeze-out regime is limited to a narrow band close to the thermalization bound.
  Quantum statistics effects are important for deriving the latter.  
  Dark matter is cold and 
  the allowed sterile neutrino   masses range from $10^{-1}$ MeV to $10^4$ TeV.  
  Such heavy neutrinos are not in conflict 
with  perturbativity as long as the dark sector is much colder than the observable one.
  
 While prospects of direct detection of  neutrino DM are rather dim,     its decay can potentially be observed for a range of tiny active-sterile mixing angles (Fig.\,\ref{limits}).
  \\ \ \\
  {\bf Acknowledgements. } We are grateful to Valentina De Romeri for collaborating at the early stages of the project and providing us with Fig.\,\ref{limits}.
  This work was supported by the JSPS Grant-in-Aid for Scientific Research
KAKENHI Grant No. JP20K22349 (TT).

\end{document}